%% file: 07_main.tex
\documentclass[sigconf]{acmart}
\usepackage{multirow}
\AtBeginDocument{%
  }

\copyrightyear{2025}
\acmYear{2025}
\setcopyright{rightsretained}
\acmConference[UMAP '25]{33rd ACM Conference on User Modeling, Adaptation and Personalization}{June 16--19, 2025}{New York City, NY, USA}
\acmBooktitle{33rd ACM Conference on User Modeling, Adaptation and Personalization (UMAP '25), June 16--19, 2025, New York City, NY, USA}
\acmDOI{10.1145/3699682.3727574}
\acmISBN{979-8-4007-1313-2/2025/06}

\begin{document}

\title{Teaming in the AI Era: AI-Augmented Frameworks for Forming, Simulating, and Optimizing Human Teams}

\author{Mohammed Almutairi}
\affiliation{Computer Sciences and Engineering\\
  \institution{University of Notre Dame}
  \city{Notre Dame}
  \state{IN}
  \country{USA}
  \postcode{46556}
}
\email{malmutai@nd.edu}

\renewcommand{\shortauthors}{Mohammed Almutairi}
%%
%% Article type: Research, Review, Discussion, Invited or position
% \acmArticleType{Position}
%%
%% Links to code and data
% \acmCodeLink{https://github.com/borisveytsman/acmart}
% \acmDataLink{htps://zenodo.org/link}
%%
%% Authors' contribution
% \acmContributions{BT and GKMT designed the study; LT, VB, and AP
  % conducted the experiments, BR, HC, CP and JS analyzed the results,
  % JPK developed analytical predictions, all authors participated in
  % writing the manuscript.}
%%
%% Sometimes the addresses are too long to fit on the page.  In this
%% case uncomments the lines below and fill them accordingly.
%%
%% \authorsaddresses{Corresponding author: Ben Trovato,
%% \href{mailto:trovato@corporation.com}{trovato@corporation.com};
%% Institute for Clarity in Documentation, P.O. Box 1212, Dublin,
%% Ohio, USA, 43017-6221}

\begin{abstract}
    \input{01_Abstract}

\end{abstract}

\keywords{Human-AI Teams, LLM-based Agent Modeling, Personalized Team Recommendation, Automated Feedback}

\begin{CCSXML}
<ccs2012>
  <concept>
    <concept_id>10003120.10003121.10003122.10003124</concept_id>
    <concept_desc>Human-centered computing~Collaborative and social computing</concept_desc>
    <concept_significance>500</concept_significance>
  </concept>
  <concept>
    <concept_id>10003120.10003121.10003122.10003125</concept_id>
    <concept_desc>Computing methodologies~Multi-agent systems</concept_desc>
    <concept_significance>300</concept_significance>
  </concept>
  <concept>
    <concept_id>10010147.10010178.10010179.10010180</concept_id>
    <concept_desc>Computing methodologies~Machine learning~Reinforcement learning</concept_desc>
    <concept_significance>300</concept_significance>
  </concept>
  <concept>
    <concept_id>10010147.10010257.10010258</concept_id>
    <concept_desc>Computing methodologies~Simulation and modeling</concept_desc>
    <concept_significance>300</concept_significance>
  </concept>
</ccs2012>
\end{CCSXML}

\ccsdesc[500]{Human-centered computing~Collaborative and social computing}
\ccsdesc[300]{Computing methodologies~Multi-agent systems}
\ccsdesc[300]{Computing methodologies~Machine learning~Reinforcement learning}
\ccsdesc[300]{Computing methodologies~Simulation and modeling}

\maketitle

\input{02_Introduction}
\input{03_Related_Work}
\input{04_Research_Objectives}
\input{05_Future_Work_Dissertation_Plan}

% \begin{acks}
% \input{06_ACKNOWLEDGMENTS}
% \end{acks}
\bibliographystyle{ACM-Reference-Format}

\bibliography{00_citation}
\end{document}

%% file: 01_Abstract.tex
Effective teamwork is essential across diverse domains. During the \textbf{team formation stage}, a key challenge is forming teams that effectively balance user preferences with task objectives to enhance overall team satisfaction. In the \textbf{team performing stage}, maintaining cohesion and engagement is critical for sustaining high team performance. However, existing computational tools and algorithms for team optimization often rely on static data inputs, narrow algorithmic objectives, or solutions tailored for specific contexts, failing to account for the dynamic interplay of team members’ personalities, evolving goals, and changing individual preferences. Therefore, teams may encounter member dissatisfaction, as purely algorithmic assignments can reduce members’ commitment to team goals or experience suboptimal engagement due to the absence of timely, personalized guidance to help members adjust their behaviors and interactions as team dynamics evolve. Ultimately, these challenges can lead to reduced overall team performance.

Driven by these challenges, my Ph.D. dissertation aims to develop AI-augmented team optimization frameworks and practical systems that enhance team satisfaction, engagement, and performance. First, I propose a team formation framework that leverages a multi-armed bandit algorithm to iteratively refine team composition based on user preferences, ensuring alignment between individual needs and collective team goals to enhance team satisfaction. Second, I introduce tAIfa (``Team AI Feedback Assistant''), an AI-powered system that utilizes large language models (LLMs) to deliver immediate, personalized feedback to both teams and individual members, enhancing cohesion and engagement. Finally, I present PuppeteerLLM, an LLM-based simulation framework that simulates multi-agent teams to model complex team dynamics within realistic environments, incorporating task-driven collaboration and long-term coordination. My work takes a human-centered approach to advance AI-driven team optimization through both theoretical frameworks and practical systems to improve team members' satisfaction, engagement, and performance.

%% file: 02_Introduction.tex
\section{INTRODUCTION}
Teams are fundamental to learning and work but often face challenges related to collaboration, engagement, and decision-making throughout their development stages. The first challenge arises in the \textbf{forming stage}, where teams must be assembled in a way that fosters consensus and satisfies all members. Traditional team formation in classrooms or workplaces is either manual or uses static assignment algorithms (e.g., matching based on skills or availability) \cite{gomez2022search}. Such approaches rarely account for individuals’ preferences or the degree of agreement among team members regarding their team composition. Research shows that involving users in team formation can boost their motivation and satisfaction \cite{alhazmi2017empirical}. Moreover, traditional recommender system techniques have been applied to suggest team members or groupings based on profiles and preferences. However, many prior approaches assume member preferences are static and often ignore the evolution of those preferences or the need to balance fairness among all members.

Once teams are formed, new challenges arise in the \textbf{performing stage}, particularly in team engagement. A major challenge in sustaining and enhancing team engagement is the lack of constructive feedback during teamwork. Without feedback aligned with team goals, members may continue to have poor communication habits or disengage, hindering overall performance \cite{jahanbakhsh2017you}. While human managers or coaches can provide guidance, their input is not always available or timely for every team interaction. Recent efforts in human-computer interaction have begun exploring AI-mediated feedback. For example, He et al. \cite{he2017two} developed a tool that analyzes team chat for signs of low participation or poor collaboration and advises members on improving their interaction norms. Similarly, Faucett et al. \cite{faucett2017should}’s ReflectLive system delivers real-time non-verbal feedback during video consultations, monitoring eye gaze and interruptions to help clinicians improve patient engagement. These works show the promise of AI in guiding teams, yet existing solutions are often limited to specific feedback types or require significant manual setup. There remains a need for a general, scalable approach to provide personalized, automated feedback that adapts to any team’s communication patterns and engagement levels.

Another key challenge in sustaining and enhancing team engagement is understanding team dynamics in complex teamwork scenarios. Researchers and practitioners often seek to experiment with different team structures, roles, or interaction strategies, but conducting such experiments with actual teams can be costly and difficult to control. Traditional simulations of team dynamics (e.g., agent-based models \cite{lapp2019kaboom}) capture some aspects of teamwork but often simplify social and communication behaviors. These limitations reduce their ability to model the nuanced, adaptive nature of human-team interaction. Recently, the emergence of LLMs has opened new possibilities for simulation. LLM agents can generate human-like behavior, suggesting they could emulate how team members discuss, plan, and collaborate on tasks \cite{park2024generative}. These advances demonstrate the potential of LLM-based simulation frameworks to create realistic, interactive environments where team interventions can be tested and refined. However, current LLM-powered simulations remain limited in their ability to model real-world complexities, such as scenarios requiring task-driven collaboration and long-term coordination \cite{gao2024large}. 

Motivated by these gaps, my Ph.D. dissertation aims to develop AI-driven frameworks and practical systems that improve team formation, engagement, and, ultimately, team performance. Specifically, my work addresses the limitations of team formation methods by designing an adaptive framework that iteratively incorporates users' preferences to improve consensus-driven team assembly. I developed a multi-armed bandit-based framework that dynamically refines team recommendations based on user feedback, ensuring alignment between individual preferences and collective team goals. Additionally, to enhance team engagement and cohesion, my research addresses the lack of real-time and personalized feedback in team interactions. I introduce \textsc{tAIfa}, an AI-mediated feedback system that analyzes team communication patterns and provides automated, context-aware recommendations to both individuals and the team as a whole. Finally, my research advances LLM-based simulations to model team dynamics in a real-world simulated environment, integrating task-driven collaboration and long-term coordination. I developed PuppeteerLLM, a simulation framework that leverages multi-agent teams powered by LLMs to emulate complex team interactions. This allows researchers to experiment with different team structures, roles, and engagement strategies in a controlled setting.

%% file: 03_Related_Work.tex
\section{RELATED WORK}
\subsection{Recommender Systems for Team Formation}
Recommender systems (RS) have been adapted to facilitate team formation by framing team assembly as a recommendation problem. Empirical studies suggest that RS-based team formation outperforms traditional methods, such as greedy approaches, by optimizing member selection based on multiple factors \cite{zhang2019recommendation, loughry2014assessing}. One widely used RS technique is collaborative filtering (CF) \cite{yang2016collaborative}, introduced by Resnick et al. \cite{resnick1994grouplens}, which recommends team members based on the preferences of similar users. Yet, CF relies on sufficient preference data to be effective and may lead to homogeneous teams, as it tends to suggest members with similar attributes. Another technique is content-based filtering (CBF) \cite{ko2022survey}, proposed by Rich \cite{rich1983users} recommends team members based on user preferences and predefined features. However, CBF effectiveness is constrained by the similarity function used to compare historical profiles with current preferences. Moreover, studies adopting this technique often assume user preferences remain static, an assumption challenged by \cite{jones2011improving}. To address the limitations of CF and CBF, the hybrid technique was introduced and has been widely used across various domains \cite{alkan2018opportunity, joshi2019team}. For example, Zhang et al. \cite{zhang2019recommendation} developed an on-demand taxi platform that uses hybrid RS techniques to form driver teams, recommending drivers based on similarity matrices. However, this method favors early leaders in the selection process, as once a leader accepts a member, that member is unavailable to others. This sequential selection reduces fairness, particularly for later leaders who must choose from a more limited pool.

To overcome the limitations of traditional RS techniques, reinforcement learning provides a dynamic approach that adapts to user preferences over time. Multi-armed bandit (MAB) algorithms, such as \cite{bubeck2012regret, thompson1933likelihood}, are suited for scenarios requiring a balance between exploring new options and exploiting known good choices. They have been applied in various adaptive decision-making problems \cite{ bastani2021mostly} and serve as a foundation for my team formation work. Zhou et al. \cite{zhou2018search} introduced a system that uses a multi-armed bandit framework to discover which team behavior policies, such as communication norms, decision-making styles, or leadership structures, work best for a particular team structure. Each potential policy is treated as an “arm” of the bandit, allowing the system to experimentally with and switch among various options in real time.

\subsection{Feedback in Teams}
Feedback interventions enhance team processes such as communication balance and coordination \cite{hackman1971employee}. Effective feedback helps teams reflect on their performance and refine their motivation when they are aware of necessary adjustments \cite{masthoff2024towards,gomez2020taxonomy}. Depending on its nature, feedback can be subjective or objective, incorporating conversational insights, behavioral patterns, or performance evaluations, and may target individuals or entire teams \cite{nadler1979effects}. Since tasks evolve over multiple cycles, leaders can monitor team members' performance, diagnose problems, and provide targeted feedback to enhance team learning and skill development \cite{kozlowski2008developing}. Research categorizes team feedback into \textit{recipient-focused}, where the feedback is directed at individuals or the whole team, and \textit{time-focused} feedback, where the feedback is delivered before, during, or after a task \cite{deeva2021review}. Real-time feedback during a task helps members stay on track, adjust their efforts toward team goals, and enhances engagement by providing immediate acknowledgment \cite{nikiforow2021contextual}. However, as teams scale in size and complexity, providing timely, personalized feedback to every team member requires substantial effort, making it difficult for leaders to effectively monitor team dynamics. AI-generated feedback offers a scalable solution to automate performance evaluation, team monitoring, and continuous improvement suggestions without human intervention \cite{daryanto2025conversate, benharrak2024writer}. Several tools have been developed to provide real-time team feedback \cite{keuning2018systematic, faucett2017should}. For example, SayWAT \cite{boyd2016saywat}, implemented on Google Glass, provides visual feedback on speech prosody. Similarly, Samrose et al. \cite{samrose2020immediate} designed a video chat platform that gives feedback on interruptions, volume, and emotional tone in team discussions. 

\subsection{Modeling and Simulation Team Dynamics}
Researchers have employed agent-based modeling (ABM) and reinforcement learning (RL) to simulate teams' behaviors and performance \cite{lapp2019kaboom}. ABMs rely on predefined rules to model agent interactions, making them effective for generalizable collective behaviors but limited in capturing complex social interactions and adaptive decision-making \cite{rand2021agent,an2021challenges}. RL-based approaches, on the other hand, enable agents to learn from their environment, allowing for more dynamic behavior modeling. For instance, Meimandi et al. \cite{meimandi2023rl} proposed an RL-based framework that integrates beliefs, prior knowledge, and social interactions to simulate ethics and trust in teams. However, RL models require extensive training and rely on fixed parameters, which can oversimplify agent decision-making and limit their ability to emulate real-world cognition \cite{zhang2021synergistic, lake2017building}. In contrast to ABM and RL, large language models (LLMs) offer a more flexible approach to simulating social interactions and evolving team behaviors \cite{gao2024large}. LLM-powered agents have demonstrated human-like decision-making and memory retention, enabling them to dynamically adapt based on prior interactions \cite{park2023generative, wu2023autogen}. However, LLM-based multi-agent systems still face challenges in coordinated multi-agent interactions to perform task-driven collaboration and long-term coordination with spatial and temporal dynamics \cite{guo2024large}. 

%% file: 04_Research_Objectives.tex
\section{RESEARCH OBJECTIVES, METHODOLOGY, AND RESULT}
My doctoral research goal is to enhance team satisfaction, engagement, and, ultimately, team performance by developing AI-driven theoretical frameworks and practical systems. To achieve this goal, I define three main research objectives, each addressing a specific set of research questions. Below, I outline these objectives and the questions they seek to answer:

\subsection{RO1: Develop a team formation framework that incorporates user preferences to maximize consensus among members in assembled teams.} 
This objective addresses the following research questions:
\begin{itemize}
    \item \textbf{RQ1.} How effective is the recommendation system in guiding users in selecting a team that aligns with their preferences?
    \item \textbf{RQ2.} How does incorporating user preferences during the team formation process impact team satisfaction?
    \item \textbf{RQ3.} How does incorporating user preferences during the team formation process impact team performance?
\end{itemize}

To address the research questions, I have structured my work into the following steps:

\subsubsection{\textbf{Algorithm Development:}} I modeled the team formation problem as a MAB problem, extending the Upper Confidence Bound (UCB) algorithm to iteratively refine team recommendations based on user feedback \cite{garivier2008upper}. My approach treats team selection as a sequential decision-making process, where the UCB algorithm adjusts recommendations by balancing exploration and exploitation. Each potential team composition is represented as an “arm,” with user feedback serving as the reward to guide the algorithm in the selection process. 

\subsubsection{\textbf{Offline Evaluation:}} I evaluated the UCB algorithm through an offline simulation, measuring how well the recommended teams align with user preferences. The evaluation leveraged Big Five personality traits as a representation of user preferences, showing that the UCB algorithm achieved a high degree of alignment between the recommended team compositions and users' selections. The preliminary result provided motivation to further examine the effect of the algorithm in real-world team interaction scenarios.

\subsubsection{\textbf{Future Work:}}
\label{RO1: Future Directions}
Building upon the preliminary results of UCB algorithm evaluation, the next step is to develop an interactive system that integrates the proposed MAB-based team formation framework within a real-time communication environment using Discord\footnote{\url{https://discord.com/developers/docs/quick-start/overview}}. Specifically, the system will allow users to specify their profiles. Then, an optimization algorithm will be employed to generate an initial set of candidate team compositions. Afterward, user preferences will iteratively guide personalized recommendations of these candidate teams. Over multiple iterations, UCB algorithm will refine and converge toward optimal team compositions based on user feedback. This iterative process will generate a preference score matrix, which will subsequently be used to solve an assignment problem to maximize overall user-team alignment. After completing system development, I will conduct a between-subjects laboratory experiment to evaluate the effectiveness of different team formation methods on team satisfaction and performance. Participants will be randomly assigned to a team, and each session will follow one of four experimental conditions: random teams, self-assembled teams, and proposed team formation framework.

\subsection{RO2: Develop an AI agent powered by LLMs to analyze team interactions and generate immediate, personalized feedback to enhance team engagement and cohesion.}

This objective addresses the following research questions:

\begin{itemize}
    \item \textbf{RQ1:} How does including AI-generated automated feedback impact team members' engagement and interactions with each other?
    \item \textbf{RQ2:} How does including AI-generated automated feedback impact team performance?
\end{itemize}

To achieve this objective and answer the research questions, I organized my work into these steps:

\subsubsection{\textbf{Prompt Design:}} I began by designing LLM prompts to generate effective feedback messages, drawing on insights from small-group research to integrate seven key communication metrics that evaluate team communication. First, language style matching (LSM) measures how closely team members align their communication styles by analyzing function word usage. Sentiment analysis captures the emotional tone of conversations, classifying them as positive, negative, or neutral. The transactive memory system assesses how well team members recognize and leverage each other’s expertise during decision-making. Team engagement measures balanced participation, ensuring that a few members do not dominate conversations. Collective pronoun usage indicates shared team identity, while communication flow examines response times, interruptions, and turn-taking patterns. Finally, topic coherence evaluates whether discussions stay aligned with the team’s objectives, reducing off-task distractions.

\subsubsection{\textbf{System Development:}} I developed tAIfa, an LLM-powered agent designed to provide immediate, AI-generated feedback to individuals and teams on Slack. The system operates through four main processing stages: Stage\textbf{\textcircled{1}} retrieves and structures team messages into a JSON format, preserving metadata and chronology; Stage\textbf{\textcircled{2}} evaluates team dynamics by computing the communication metrics; Stage\textbf{\textcircled{3}} uses these metrics and structured prompts to generate personalized feedback through the LLM; and Stage\textbf{\textcircled{4}} delivers both individual feedback via private message and team feedback via public message. The feedback is structured into four key sections: task contribution summary, strengths, areas for improvement, and actionable recommendations.

\subsubsection{\textbf{User Study:}}
To evaluate the effectiveness of tAIfa, I conducted a laboratory experiment to examine its impact on team engagement, cohesion, and performance. A total of 54 participants were assigned to teams of three and performed three collaborative tasks, each lasting 10 minutes, simulating real-world team dynamics on Slack. This between-subjects study included a control group (no AI feedback) and a treatment group (receiving AI feedback from tAIfa). Each 60-minute session followed a three-round structure, after which participants completed surveys on their team experiences, interactions, and engagement levels.

\subsubsection{\textbf{Results:}} My preliminary results indicate that teams in the treatment condition outperformed the control group across multiple engagement metrics. They engaged in deeper discussions, had longer conversations, and involved more participants, suggesting that tAIfa feedback fostered active participation and sustained engagement. Participants in the treatment group also reported higher levels of team efficiency, satisfaction, and willingness to collaborate with the same team again. Additionally, teams in the treatment condition showed an improvement in overall performance compared to the control teams.

\begin{table}[h]
    \centering
    \renewcommand{\arraystretch}{0.5}
    \setlength{\tabcolsep}{1pt}
    \begin{tabular}{lcc}
        \hline
        \textbf{Metric (Average)} & \textbf{Control} & \textbf{Treatment} \\
        \hline
        Conversation Duration (Minutes) & 6.9 & 8.14 \\
        Speaker Turn Frequency & 16 & 20.87 \\
        Word Count & 233.04 & 260.3 \\
        Task Performance & 60.4\% & 62\% \\
        \hline
    \end{tabular}
    \label{tab:avg_engagement_metrics}
\end{table}

\subsubsection{\textbf{Future Work:}} 
\label{RO2: Future Directions}
As of now, the tAIfa agent does not support real-time users interaction and the feedback is provided only at the end of the tasks. I plan to expand the agent’s capabilities to enable dynamic interaction, allowing users to choose their preferred feedback timing. Also, I aim to recruit larger and more diverse samples to ensure statistically significant findings.

\subsection{RO3: Develop an LLM-based simulation framework to study team dynamics.}

This objective addresses the following research question:
\begin{itemize}
    \item \textbf{RQ1:} How can LLMs be leveraged to create simulation environments that accurately emulate human-team interactions?
\end{itemize}

To achieve this objective and answer the research question, I organized my work into these steps:

\subsubsection{\textbf{Framework Development:}} 
I developed PuppeteerLLM, a simulation framework that models complex human team dynamics by incorporating LLM agents, physical environments (allowing agents to “move” and occupy locations affecting their interactions), and temporal dynamics (progressing over time with agents remembering and reacting to past events). The framework proceeds through three stages: \textit{Initialization}, where the system sets up the environment and agents; \textit{Execution}, where LLM agents iteratively generate and validate events; and \textit{Termination}, where the simulation ends and the final state is recorded. PuppeteerLLM represents the environment as a graph-based structure \(\mathcal{G} = \{ g_1, g_2, \dots, g_n \}\) where each leaf node corresponds to a unique spatial region, allowing agents to navigate and reason about physical constraints. The framework provides agents with conversation capability (dyadic and nested conversations) and an event scheduling system that queues each event and executes it in a coherent timeline, thereby mitigating challenges like LLM hallucinations. Finally, PuppeteerLLM generates a structured JSON log of each simulation, supporting both quantitative and qualitative analysis of agent behaviors and team dynamics.

\subsubsection{\textbf{Future Work:}} 
\label{RO3: Future Directions}
My next step is to develop a user interface that assists users in setting up the scenario, configuring the agents, and customizing the environment. This interface will allow researchers to create tasks, define team roles, personalities, and environments without direct code manipulation. Upon the completion of the user interface, I plan to conduct two experiments. First, a comparative analysis will compare simulation outputs to a ground truth team dataset, assessing whether the framework replicates human-like behaviors within the same scenario. This step requires defining which metrics resemble team dynamics to compare. Second, I will conduct a user study in which participants interact with the PuppeteerLLM interface to simulate team dynamics. Then, I will survey their perceptions of the simulated team interaction. 

%% file: 05_Future_Work_Dissertation_Plan.tex
\section{DISSERTATION STATUS AND LONG-TERM PLAN}
I started my Ph.D. \textit{in January 2023} in Computer Science and Engineering at the University of Notre Dame under the supervision of professor \href{https://www.dgomezara.cl/}{Diego Gómez-Zará}. Currently, I am in my \textbf{third year of study}, with a projected completion date in \textbf{Fall 2026}. Over the next two years, I will pursue the future work outlined in Sections \ref{RO1: Future Directions}, \ref{RO2: Future Directions}, and \ref{RO3: Future Directions} to address my research objectives. In the long term, I plan to focus on advancing LLM-based agent simulations to model complex team dynamics and provide researchers with tools to enhance team performance. The overall goal of my thesis is to develop AI-driven frameworks and practical systems that enhance team formation, engagement, and cohesion across different stages of team development.